\documentclass[aps,pra,reprint,amsmath,amssymb,superscriptaddress,nobibnotes]{revtex4-1}
\usepackage{ulem,bm}
\usepackage{amsmath, upgreek, amssymb, graphicx, paralist, gensymb,epstopdf}
\usepackage[colorlinks, linkcolor=blue, citecolor=blue, urlcolor=blue]{hyperref}
\usepackage{array,epsfig,amsmath,amssymb}
\usepackage{graphicx}
\usepackage{color}

\begin{document}

\title{Implementation of a $3\times3$ directionally-unbiased linear optical multiport}

\author{Ilhwan Kim}
\affiliation{Center for Quantum Information, Korea Institute of Science and Technology (KIST), Seoul, 02792, Korea}
\affiliation{Department of Applied Physics, Institute of Natural Science, Kyung Hee University, Yongin-si 17104, Korea}

\author{Donghwa Lee}
\affiliation{Center for Quantum Information, Korea Institute of Science and Technology (KIST), Seoul, 02792, Korea}
\affiliation{Division of Nano and Information Technology, KIST School, Korea University of Science and Technology, Seoul 02792, Korea}

\author{Seongjin Hong}
\affiliation{Center for Quantum Information, Korea Institute of Science and Technology (KIST), Seoul, 02792, Korea}

\author{Young-Wook Cho}
\affiliation{Center for Quantum Information, Korea Institute of Science and Technology (KIST), Seoul, 02792, Korea}
\affiliation{Department of Physics, Yonsei University, Seoul 03722, Korea}

\author{Kwang Jo Lee}
\affiliation{Department of Applied Physics, Institute of Natural Science, Kyung Hee University, Yongin-si 17104, Korea}

\author{Yong-Su Kim}
\email{yong-su.kim@kist.re.kr}
\affiliation{Center for Quantum Information, Korea Institute of Science and Technology (KIST), Seoul, 02792, Korea}
\affiliation{Division of Nano and Information Technology, KIST School, Korea University of Science and Technology, Seoul 02792, Korea}

\author{Hyang-Tag Lim}
\email{hyangtag.lim@kist.re.kr}
\affiliation{Center for Quantum Information, Korea Institute of Science and Technology (KIST), Seoul, 02792, Korea}
\affiliation{Division of Nano and Information Technology, KIST School, Korea University of Science and Technology, Seoul 02792, Korea}

\date{\today} 

\begin{abstract}
Linear optical multiports are widely used in photonic quantum information processing. Naturally, these devices are directionally-biased since photons always propagate from the input ports toward the output ports. Recently, the concept of directionally-unbiased linear optical multiports was proposed. These directionally-unbiased multiports  allow photons to propagate along a reverse direction, which can greatly reduce the number of required linear optical elements for complicated linear optical quantum networks. Here, we report an experimental demonstration of a $3 \times 3$ directionally-unbiased linear optical fiber multiport using an optical tritter and mirrors. Compared to the previous demonstration using bulk optical elements which works only with light sources with a long coherence length, our experimental directionally-unbiased $3 \times 3$ optical multiport does not require a long coherence length since it provides negligible optical path length differences among all possible optical trajectories. It can be a useful building block for implementing large-scale quantum walks on complex graph networks. 
\end{abstract}

\maketitle

\section{Introduction}
	
Quantum walks (QW), the quantum analogue of classical random walks, provide promising methodologies to execute complex quantum algorithms for quantum computing~\cite{Kempe2003,Venegas2012}. As random walks are used for algorithm design in classical computer, many quantum algorithms are developed based on QW which provide speed-up over the best known classical algorithms for the same problem~\cite{Childs2002}. Furthermore, this approach has capability of performing algorithms for universal quantum computation~\cite{Crespi2013,Simon2018,Childs2009,Aharonov1993,Aharonov2001}. Linear optics has been considered as one of the most promising candidates for implementing QW, because of its robustness against decoherence, ease of preparation of quantum states and its scalability through on-chip integration~\cite{Carolan2015,Carolan2014,Tillmann2013,Tang2018}.

QW are movements of walkers on a graph, which consist of connected vertices with \textit{N} edges~\cite{Tang2018_2, Faccin2013, Tregenna2003, Bian2017}. A walker moves from a vertex to one of the neighboring vertices at each step. In linear optics, each vertex can be implemented by an $\textit{N} \times \textit{N}$ multiport which is a linear optical element having \textit{N} input ports and $N$ output ports. A set of phase shifters and $2 \times 2$ beam splitters (BSs) can construct an $N \times N$ multiport with $N\geq3$~\cite{Reck1994,Clements2016}. Since directionality of a BS forbids input photons to move back to the same port where the input photons entered, an implementation of an $N \times N$ multiport forms a unidirectional expending tree-like structure to cover all possible optical paths. From the inability of reverse propagation, this unidirectional implementation is called \textit{a directionally-biased multiport}.
	
Recently, \textit{a directionally-unbiased multiport} has been proposed~\cite{Simon2016}. Unlike a directionally-biased multiport, it allows an input photon to leave the same port where the input photon entered. Hence, it is suitable for implementing undirected graphs for QW. The bidirectional feature of the directionally-unbiased multiports can surprisingly reduce the resource requirement for some QW configurations \cite{Simon2016}.

It has been reported only an experimental implementation of an $N \times N$ directionally-unbiased multiport for $N\geq3$~\cite{Osawa2018}. Osawa \textit{et al.} implemented a $3 \times 3$ directionally-unbiased multiport based on an optical cavity in free space using bulk optical elements~\cite{Osawa2018}. However, this scheme requires input photons having a long coherence length due to large path length differences among the different trajectories in the cavity. Indeed, it was experimentally tested with a HeNe laser beam whose coherence length is longer than 1~km. It does not show the quantum optical phenemoena in directionally-unbiased multiports, and yet there has been no experimental demonstration of directionally-unbiased multiport using quantum linght sources.	
		
In this paper, we experimentally demonstrate a $3 \times 3$ directionally-unbiased linear optical multiport using a $3 \times 3$ optical fiber BS, or a tritter. By measuring the two-photon Hong-Ou-Mandel (HOM) interference~\cite{Hong1987} and amplitude distribution ratios of the $3 \times 3$ directionally-unbiased multiport, we reconstructed a transfer matrix of the implemented multiport. All results are obtained using quantum light sources. We believe that our directionally-unbiased multiport device can be useful building blocks for simulating QWs on complex graph networks.

\section{Theory}
	
\subsection{Directionally-biased/unbiased multiports using fiber optical multiports}

\begin{figure}[t]
\centering\includegraphics[width=3.4 in]{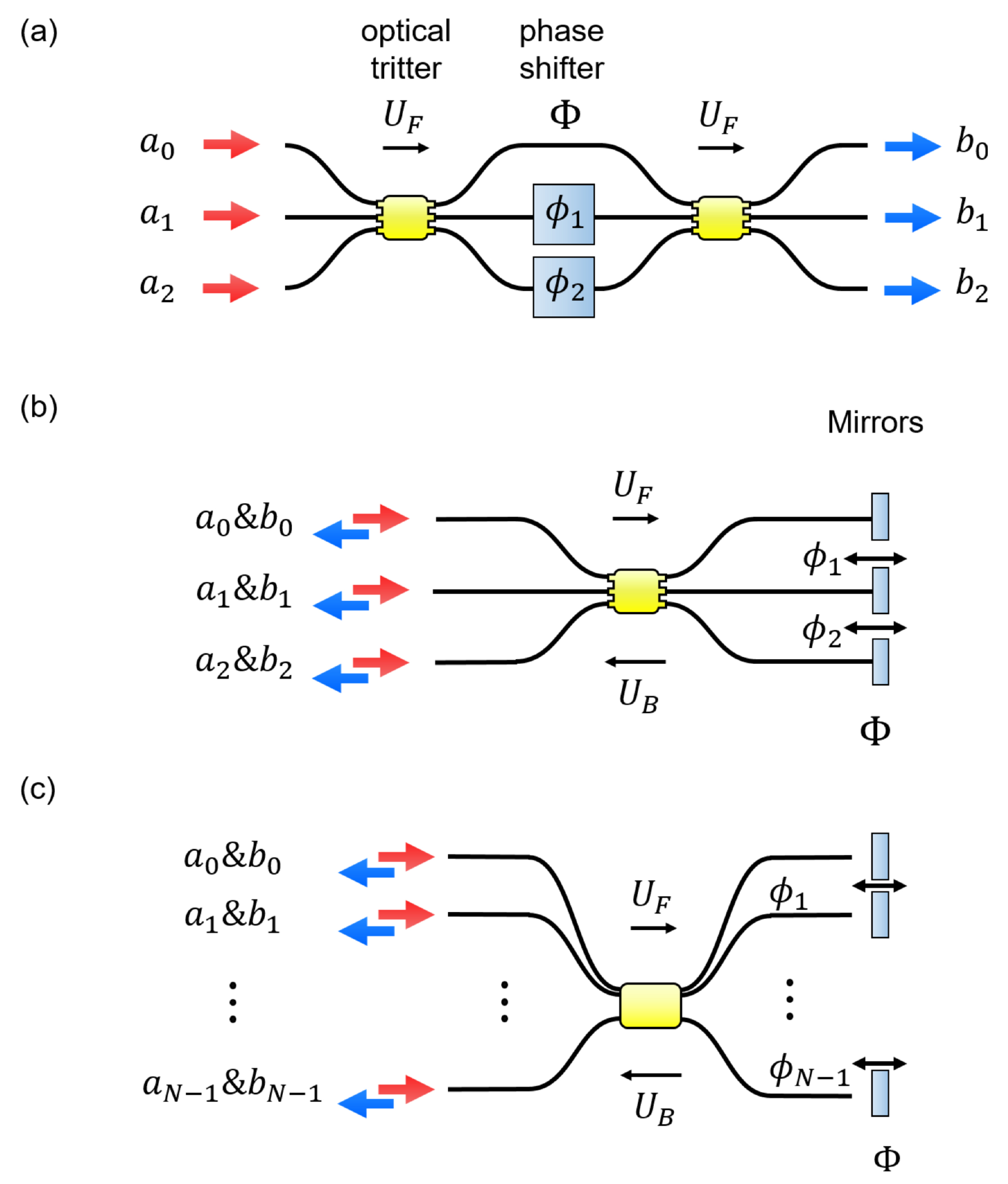}
\caption{Schematic designs of (a) a directionally-biased $3 \times 3$ optical multiport (b) a directionally-unbiased $3 \times 3$ optical multiport, and (c) a directionally-unbiased $N \times N$ optical multiport. The relative phase of the mode 0, $\phi_{0}$, is set to zero as a reference. A directionally-biased mulitport consists of two tritters and two phase shifters $\phi_{1}$ and $\phi_{2}$, which can control the amplitude distribution of the multiport. Photons are inserted to the input ports $a_0$, $a_1$ and $a_2$, and then come out from the output ports $b_0$, $b_1$ and $b_2$. On the other hand, a directionally-unbiased multiport consists of one tritter and three mirror units. In this design, the input photons can move back to the same input ports $a_0 \&b_0$, $a_1 \&b_1$, and $a_2 \&b_2$, which means that the input and output ports are not separated. A directionally-unbiased $N\times N$ optimal multiport can be implemented by replacing a tritter with an $N\times N$ multiport and adding a mirror in each mode. The red and blue arrows correspond to the directions of the input and the output modes, respectively.}
\label{fig1}
\end{figure}

Let us consider a $3 \times 3$ linear optical multiport as shown in Fig.~\ref{fig1}(a) and (b) where a tritter is a $3 \times 3$ BS based on a fused optical fiber ~\cite{Osawa2019}. These multiports form an interferometer to control an amplitude distribution among input and output modes using phase shifters. Fig.~\ref{fig1}(a) describes a directionally-biased $3 \times 3$ linear optical multiport. This model, an expanded Mach-Zehnder interferometer, consists of two tritters and two phase shifters. Then, the transfer matrix of the $3 \times 3$ directionally-biased linear optical multiport is given by
\begin{equation}
U_{\text{biased}}=U_{F} \varPhi U_{F},
\label{eq1}
\end{equation}
where $U_{F}$ is a transfer matrix of a forward directional tritter and $\varPhi$ is a phase shift operation. For example, the transfer matrix of an ideal tritter is given by
\begin{equation}
U_{F}^{\text{ideal}}=\frac{1}{\sqrt{3}}\left(\begin{array}{ccc} 1 & 1 & 1\\1 & e^{i 2\pi/3} & e^{-i 2\pi/3}\\1 & e^{-i 2\pi/3} & e^{i 2\pi/3} \end{array}\right).
\label{eq2}
\end{equation}
The phase shift operation $\varPhi$ can be represented by the following unitary phase matrix 
\begin{equation}
\varPhi=\left(\begin{array}{ccc} 1 & 0 & 0\\0 & \textit{e}^{i\phi_{1}} & 0\\0 & 0 & \textit{e}^{i\phi_{2}} \end{array}\right),
\label{eq3}
\end{equation}
where the relative phase of the mode 0 is fixed to $\phi_0=0$ as a reference. 
	
On the other hand, a directionally-unbiased $3 \times 3$ linear optical multiport is shown in Fig.~\ref{fig1}(b). This scheme can be considered as an expanded Michelson interferometer, where the input and the output ports are the same. It can be realized by placing a mirror at each output mode of a tritter, and then the photons coming out from the output modes of a tritter is reflected back and enter the output modes of a tritter. The relative phase shift at each mode can be controlled by adjusting the position of mirrors. Note that the relative phase shift $\phi_{1}$ and $\phi_{2}$ provide the tunability on the transfer matrix of a directionally-unbiased multiport.

The transfer matrix of a directionally-unbiased $3 \times 3$ multiport $U_{\text{unbiased}}$ is similar to a directionally-biased one $U_{\text{biased}}$ except the fact that an operation of the second tritter corresponds to the backward propagation of the first one. Therefore, the overall transfer matrix is given by
\begin{equation}
U_{\text{unbiased}}=U_{B} \varPhi U_{F},
\label{eq4}
\end{equation}
where $U_{B}$ is a transfer matrix of a tritter in a backward direction, and it is equivalent to a transpose of a tritter operation in a forward direction (i.e. $U_{B}=U_{F}^{T}$). As shown in Fig.~\ref{fig1}(c), a directionally-unbiased multiport can be extended to an $N \times N$ multiport.

\subsection{Reconstruction of a transfer matrix}
	
Here we discuss the experimental reconstruction method of a transfer matrix of a general $3 \times 3$ linear optical multiport, which has three input ports \{$a_{0}$, $a_{1}$, $a_{2}$\}  and three output ports \{$b_{0}$, $b_{1}$, $b_{2}$\}. It is known that an unknown $3 \times 3$ linear optical multiport can be reconstructed by measuring the classical amplitude distribution of the multiport and the two-photon HOM quantum interference~\cite{Peruzzo2011,Spagnolo2013}. 

An unknown $3 \times 3$ unitary matrix $U$ can be represented by
\begin{equation}
U=\left(\begin{array}{ccc} |u_{00}| & |u_{01}| & |u_{02}|\\|u_{10}| & |u_{11}|e^{i\phi_{\alpha}} & |u_{12}|e^{i\phi_{\beta}}\\|u_{20}| & |u_{21}|e^{i\phi_{\gamma}} & |u_{22}|e^{i\phi_{\delta}} \end{array}\right),
\label{eq5}
\end{equation}
where $|u_{ki}|^{2}$ denotes the classical amplitude distribution of the single photon counts at the output port $b_{k}$ when single photons are inserted at the input port $a_{i}$, and $\phi_{\alpha}$, $\phi_{\beta}$, $\phi_{\gamma}$, and $\phi_{\delta}$ denote the relative phases. First, we measure the classical amplitude distribution by measuring the single photon counts $n_{ki}$ as follows:
\begin{equation}
|u_{ki}|^{2}=\frac{n_{ki}}{\varSigma^{2}_{k=0} n_{ki}},
\label{eq6}
\end{equation}
where $n_{ki}$ denotes the single photon counts at the output port $b_{k}$ when single photons are inserted at the input port $a_{i}$ ($i,k=0,1,2$). Then, we measure the visibility of the two-photon HOM quantum interference with the two-photon input state $\hat{a}^{\dagger}_{i}\hat{a}^{\dagger}_{j}|0\rangle$ where $i,j=0,1,2~(i\neq j)$. When two distinguishable photons are injected to the input ports of \textit{i} and \textit{j}, the probability $C_{ij}^{kl}$ to detect photon at each $k$ and $l$ $(k\neq l)$ output modes becomes
\begin{equation}
C_{ij}^{kl}=|U_{ki}U_{lj}|^{2}+|U_{li}U_{kj}|^{2},
\label{eq7}
\end{equation}
where $U_{ki}$ denotes the transition amplitude of the photon entering in the input port \textit{i} and exiting from output port \textit{k}. No interference occurs between two terms since two input photons are distinguishable. In the case of indistinguishable photons, the probability $Q_{ij}^{kl}$ to detect a photon at each $k$ and $l$ $(k\neq l)$ output mode for photons injected to the input ports of \textit{i} and \textit{j} ($i\neq j$) is given by
\begin{equation}
Q_{ij}^{kl}=|U_{ki}U_{lj}+U_{li}U_{kj}|^{2}.
\label{eq8}
\end{equation}
By varying the relative time delay between two input photons, we can observe the HOM interference that leads to peaks or dips in the output coincidence events when the relative time delay is zero. The HOM visibility between two input photons is determined by the phase values in the matrix $U$ and is given by
	\begin{equation}
	\mathcal{V}_{ij}^{kl}=\frac{C_{ij}^{kl}-Q_{ij}^{kl}}{C_{ij}^{kl}}.
	\label{eq9}
	\end{equation}
	Note that a positive (negative) value of $\mathcal{V}_{ij}^{kl}$ corresponds to a dip (peak).

	From the amplitude distribution and the HOM visibility results, an unknown transfer matrix $U$ can be numerically reconstructed by minimizing the root mean square distance between two visibility matrices, which is given by
	\begin{equation}
	RMS=\sum(\mathcal{V}_{U}-\mathcal{V}_{M})^{2},
	\label{eq10}
	\end{equation}
	where $\mathcal{V}_{M}$ is an experimentally measured visibility matrix and $\mathcal{V}_{U}$ is the visibility matrix calculated from the reconstructed transition matrix $U$. Numerical optimization consists of two steps. At first, relative phases $\phi_\alpha$, $\phi_\beta$, $\phi_\gamma$, and $\phi_\delta$ vary while the probability distribution ratio is fixed. The fixed distribution ratio is obtained from the measured amplitude distribution ratio $|u_{ki}|^{2}$. This step is for determining initial relative phase values. Secondly, both phases and probability distribution ratios vary for optimization.
	
	We quantified the quality of reconstruction using the similarity $S$, which is defined as \cite{Spagnolo2013}
	\begin{equation}
	S=1-\sum|\mathcal{V}_{U}-\mathcal{V}_{M}|/18,
	\label{eq11}
	\end{equation}
	where the similarity can be $0\le S\le1$. Note that $S=0$ and $1$ correspond to a completely anti-correlated and perfect overlap, respectively.

\section{Experiment}
\begin{figure*}[t]
\centering
\includegraphics[width=6.5in]{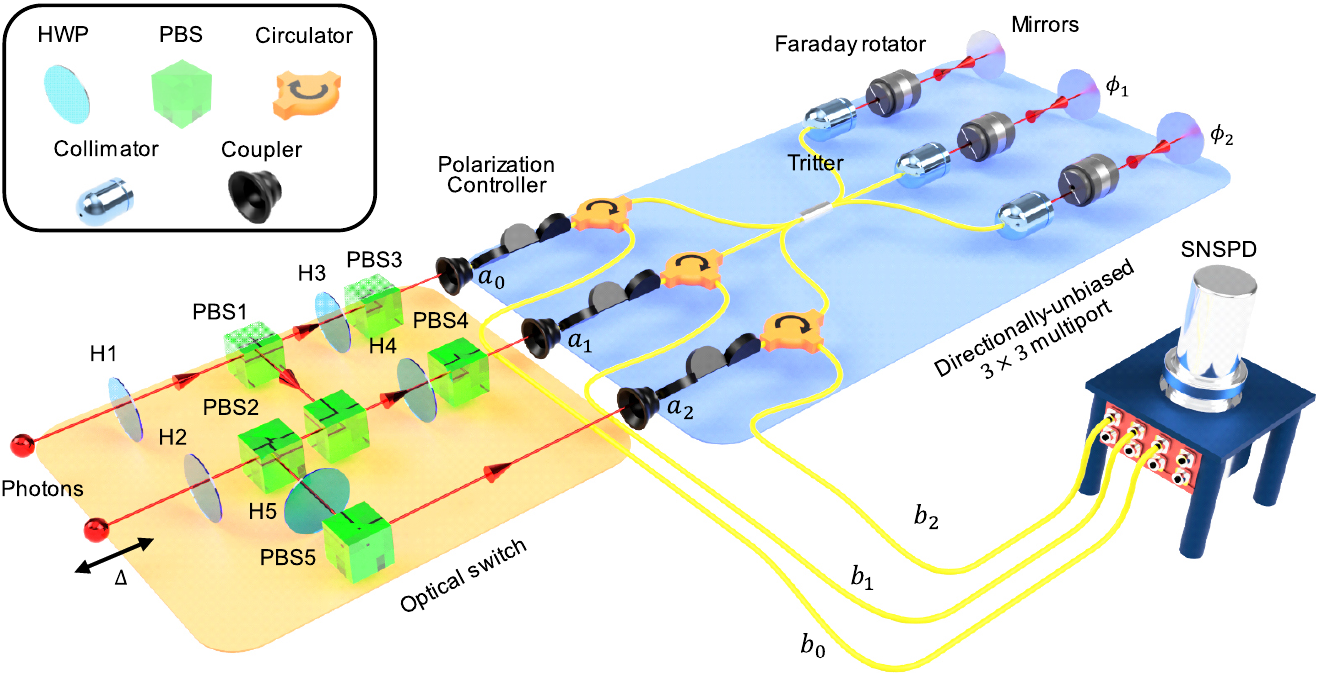}
\caption{Experimental setup for implementing a directionally-unbiased $3 \times 3$ multiport. Whole system is enclosed by an outer cage for minimizing detrimental noise from external environment so that the relative phases $\Phi$ is maintained. In order to confirm that the $\Phi$ is maintained during the experiment, we have checked that the amplitude distribution ratios are maintained during the experiment. HWP: half-waveplate, PBS: polarizing beam splitter, SNSPD: superconducting nanowire single photon detectors.}
\label{fig2}
\end{figure*}
	The schematic of our experimental setup to implement and analyze a $3 \times 3$ directionally-unbiased optical fiber multiport is shown in Fig.~\ref{fig2}. We generate a pair of identical single photons with the center wavelength of 1556~nm via type-II spontaneous parametric down conversion (SPDC) process using a 10 mm-thick PPKTP crystal (poling period of 46.15 $\mu$m) pumped by femtosecond laser pulses (not shown in Fig.~\ref{fig2}). The power and repetition rate of the pump are 10 mW and 125 MHz, respectively. The indistinguishability between photons is verified by the HOM interference visibility of $\mathcal{V} \sim 0.99$~\cite{Lee2021}. Note that the coherence length of our down-converted photons is $\sim 440 \mu$m, which is much narrower than the coherence length of a He-Ne laser.
	
	By using a set of waveplates and polarizing beam splitters (PBSs), we can prepare the input state $\hat{a}^{\dagger}_{i}\hat{a}^{\dagger}_{j}|0\rangle$ and $\hat{a}^{\dagger}_{i}|0\rangle$ with horizontal polarization where $i,j=0,1,2~(i\neq j)$ for measuring visibility of the two-photon HOM interference and amplitude distributions, respectively. In order to choose the combination of the input ports, we adjust the polarization of the input photons using a set of half waveplates (Hs) as shown in the optical switch of Fig.~\ref{fig2}. For example, $\hat{a}^{\dagger}_{0}\hat{a}^{\dagger}_{1}|0\rangle$ input state can be prepared by setting the angles of (H1, H2, H3, H4, H5) as (0$^{\circ}$, 0$^{\circ}$, 0$^{\circ}$, 0$^{\circ}$, 45$^{\circ}$). Additionally, we can obtain either $\hat{a}^{\dagger}_{0}|0\rangle$ or $\hat{a}^{\dagger}_{1}|0\rangle$ input state by setting the angles of (H3, H4) as either (0$^{\circ}$, 45$^{\circ}$) or (45$^{\circ}$, 0$^{\circ}$), respectively. Using this method, we can send all combinations of single and two-photon input states into our  $3 \times 3$ multiport.
	
	Note that photons at each input mode should have the same polarization for indistinguishability between two photons. We can adjust the polarization of the photons inside a tritter using fiber polarization controllers. Moreover, Faraday rotators located between the output mode of a tritter and a mirror guarantee that photons propagating in the opposite directions do not interfere with each other.  We can isolate the photons coming out from the input ports using optical fiber circulators, and send them toward superconducting nanowire single photon detectors (SNSPDs) for single photon detection. Single photon count rates for $\hat{b}^{\dagger}_{k}|0\rangle$ with input states $\hat{a}^{\dagger}_{i}|0\rangle$ are obtained for measuring amplitude distribution ratios, and all possible combinations of the two-photon coincidence detection events $\hat{b}^{\dagger}_{k}\hat{b}^{\dagger}_{l}|0\rangle$ where $k,l=0,1,2~(k\neq l)$ with input states $\hat{a}^{\dagger}_{i}\hat{a}^{\dagger}_{j}|0\rangle~(i\neq j)$ for measuring the two-photon HOM interference are obtained using a home-made coincidence counting unit (CCU)~\cite{Park2015,Park2021}.
	
	For measuring visibilities of the two-photon HOM interference, we introduced an optical delay $\Delta$ between two input photons by scanning the position of one collimator using a motorized linear stage. We measure all possible combinations of the two-photon HOM interference fringe for each input state $\hat{a}^{\dagger}_{i}\hat{a}^{\dagger}_{j}|0\rangle$. In order to obtain amplitude distribution ratios of the multiport, all possible single photon count rates $n_{ki}$ are measured for each input state $\hat{a}^{\dagger}_{i}|0\rangle$. The position of the motorized linear stage is rearranged whenever the input state is changed, to set an optical delay $\Delta=0$. 
	
\begin{figure}[t!]
\centering
\includegraphics[width=3.4 in]{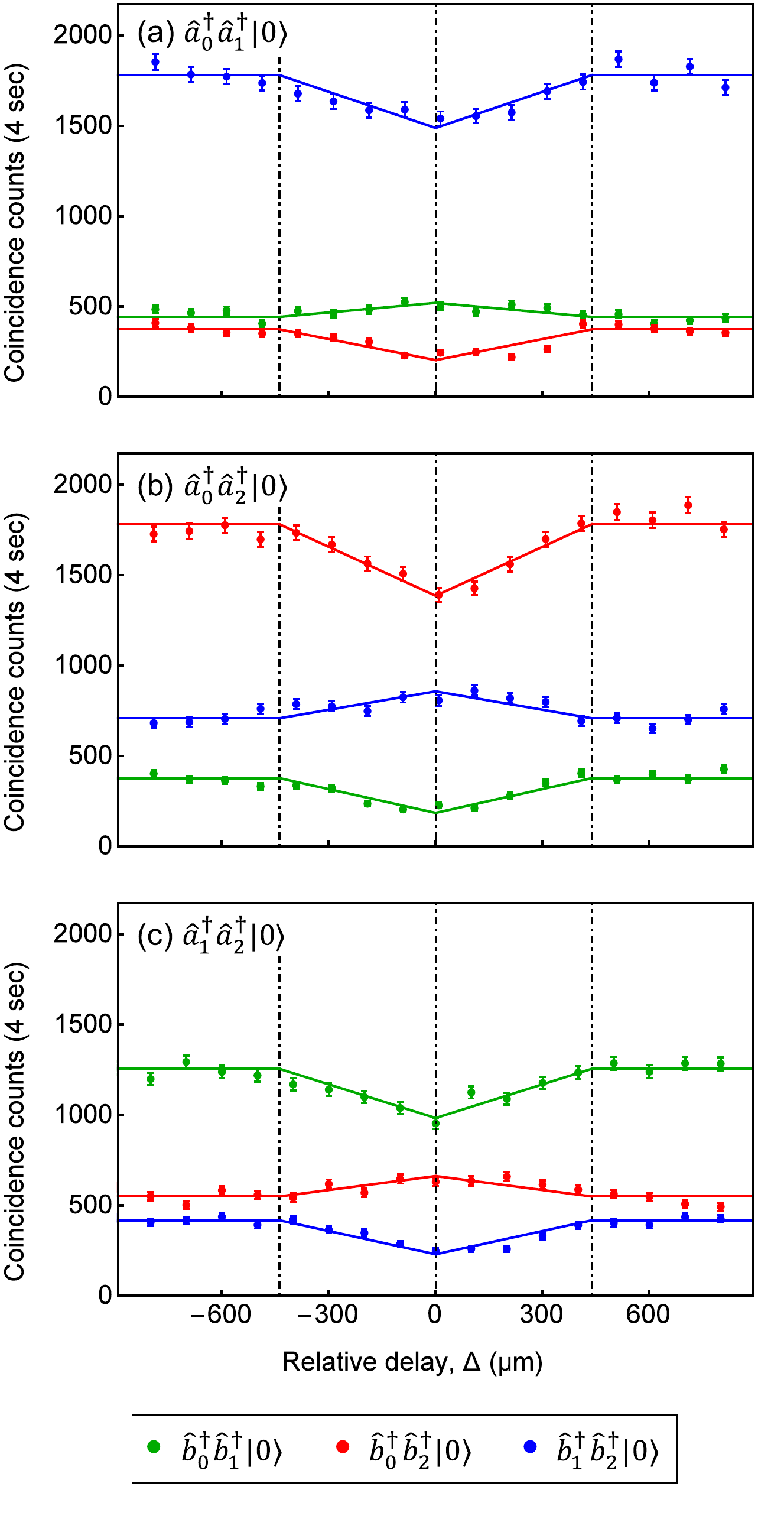}
\caption{Two-photon HOM interference results for input states (a) $\hat{a}^{\dagger}_{0}\hat{a}^{\dagger}_{1}|0\rangle$, (b) $\hat{a}^{\dagger}_{0}\hat{a}^{\dagger}_{2}|0\rangle$, and (c) $\hat{a}^{\dagger}_{1}\hat{a}^{\dagger}_{2}|0\rangle$. Green, red, and blue dots correspond to the experimental two-photon coincidence counts with detection basis $\hat{b}^{\dagger}_{0}\hat{b}^{\dagger}_{1}|0\rangle$, $ \hat{b}^{\dagger}_{0}\hat{b}^{\dagger}_{2}|0\rangle$, and $\hat{b}^{\dagger}_{1}\hat{b}^{\dagger}_{2}|0\rangle$, respectively. Solid lines are fitting functions and we calculate the visibility of the HOM interference from the fitting results. Error bars correspond to one standard deviation.}
\label{fig3}
\end{figure}

\section{Experimental results}
		
Here, we experimentally reconstruct a transfer matrix of our implemented directionally-unbiased multiport devices as shown in Fig.~\ref{fig2} using two independent reconstruction methods. First, we directly reconstruct a transfer matrix $V$ as Eq. (\ref{eq5}) by minimizing Eq. (\ref{eq10}). Second, we reconstruct the transfer matrix $W=U_{B} \varPhi U_{F}$ as Eq. (4), and by independently reconstructing $U_{F}$ and $U_{B}$, and $\varPhi$.  Then, we compare two reconstructed transfer matrices $V$ and $W$.
			 
We first experimentally reconstruct the optimal transfer matrix $V$ of our $3 \times 3$ directionally-unbiased multiport using the measured HOM visibility matrix $\mathcal{V}_{M}$ and the amplitude distribution $|u_{ki}|^{2}$. All possible combinations of the HOM interferences are measured, and experimental results and corresponding fitting functions are shown in Fig.~\ref{fig3}. From the experimental results, $\mathcal{V}_M$ is obtained to be
	
\begin{equation}
\mathcal{V}_{M}=\left(\begin{array}{ccc} -0.176\pm0.061 & 0.455\pm0.067 & 0.164\pm0.015\\0.509\pm0.083 & 0.222\pm0.018 & -0.210\pm0.049\\0.217\pm0.019 & -0.06\pm0.046 & 0.450\pm0.055 \end{array}\right),
\label{eq12}
\end{equation}
where the rows \textit{x} and columns \textit{y} of the matrix $\mathcal{V}_{M}$ correspond to the input state  $x=\{ \hat{a}^{\dagger}_{0}\hat{a}^{\dagger}_{1}|0\rangle,  \hat{a}^{\dagger}_{0}\hat{a}^{\dagger}_{2}|0\rangle,  \hat{a}^{\dagger}_{1}\hat{a}^{\dagger}_{2}|0\rangle \}$ and the output detection basis $y=\{ \hat{b}^{\dagger}_{0}\hat{b}^{\dagger}_{1}|0\rangle,  \hat{b}^{\dagger}_{0}\hat{b}^{\dagger}_{2}|0\rangle,  \hat{b}^{\dagger}_{1}\hat{b}^{\dagger}_{2}|0\rangle \}$, respectively.
	
From the measured count rate, the amplitude distribution $|u_{M}|^{2}$ is obtained by using Eq. (\ref{eq6}). The matrix of the amplitude distribution $|u_{M}|^{2}$ is obtained to be
\begin{equation}
|u_{M}|^{2}=\left(\begin{array}{ccc} 0.220\pm0.001 & 0.131\pm0.001 & 0.648\pm0.002\\0.150\pm0.001 & 0.610\pm0.002 & 0.241\pm0.001\\0.679\pm0.001 & 0.190\pm0.001 & 0.131\pm0.001 \end{array}\right).
\label{eq13}
\end{equation}
From the measurement results $\mathcal{V}_{M}$ and $|u_{M}|^{2}$, we numerically found a transfer matrix $V$ of our implemented directionally-unbiased $3 \times 3$ linear optical multiport. The experimentally reconstructed matrix $V$ and the corresponding standard deviation $\sigma_{V}$ are given by
	
\begin{figure}[t]
\centering
\includegraphics[width=3.3 in]{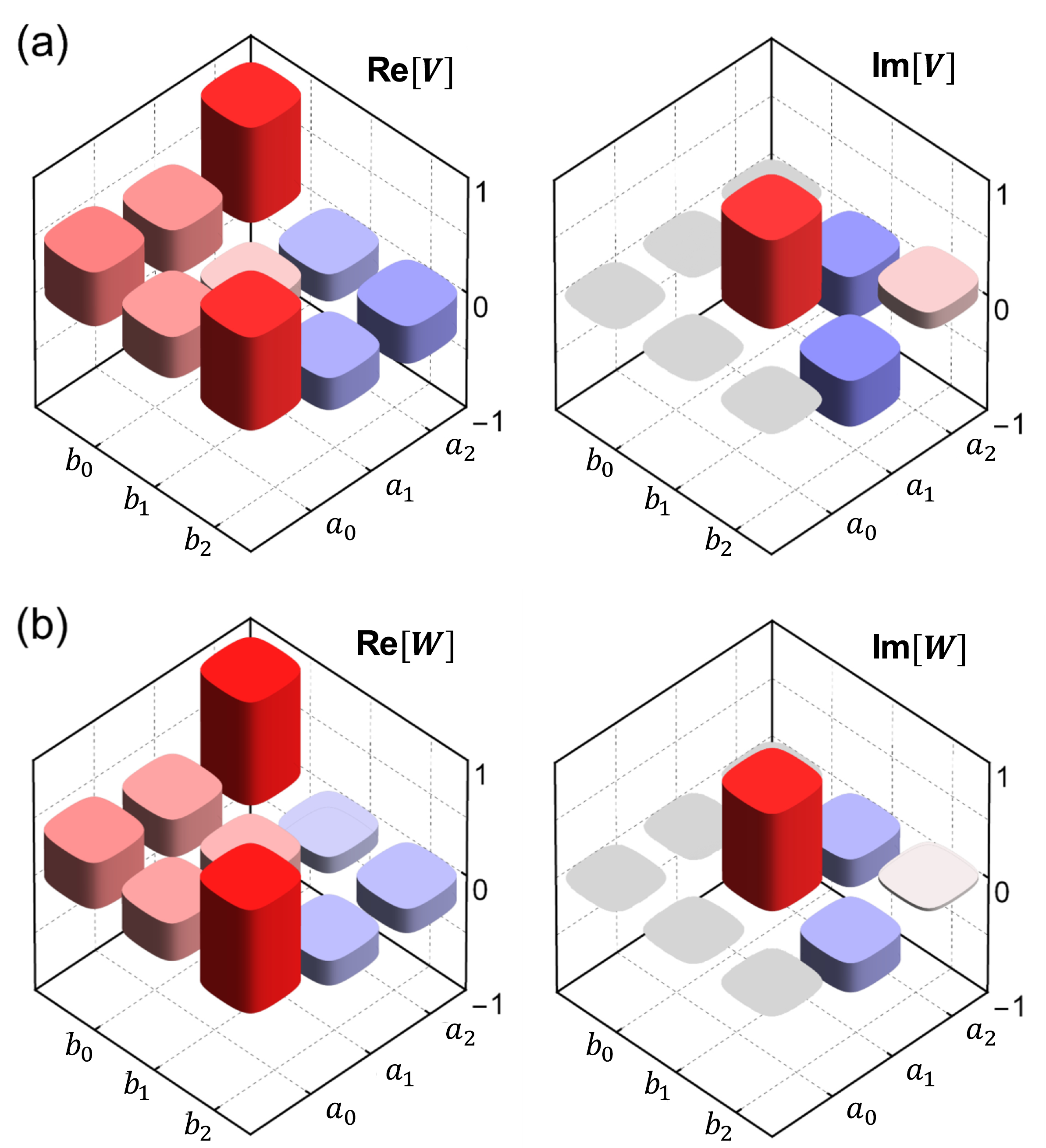}
\caption{Real and imaginary parts of the experimentally reconstructed transfer matrices of (a) $V$ obtained by optimizing Eq. (\ref{eq10}) and (b) $W$ obtained by reconstructing each operation in Eq. (\ref{eq4}). The fidelity between them is calculated as $F(V, W)=0.971\pm0.005$.}
\label{fig4}
\end{figure}
	
\begin{equation}
V=\left(\begin{array}{ccc} 0.466 & 0.358 & 0.807 \\ 0.382 & 0.783 \, e^{i \, 0.437\pi} & 0.487 \, e^{-i \, 0.694\pi} \\ 0.826 & 0.433 \, e^{-i \, 0.682\pi} & 0.357 \, e^{i \, 0.870\pi} \end{array}\right),
\label{eq14}
\end{equation}
and
\begin{equation}
\sigma_{V}=\left(\begin{array}{ccc} 0.001 & 0.002 & 0.001 \\ 0.002 & 0.001 \, e^{i \, 0.028\pi} & 0.001 \, e^{i \, 0.031\pi} \\ 0.001 & 0.001 \, e^{i \, 0.027\pi} & 0.001 \, e^{i \, 0.096\pi} \end{array}\right),
\label{eq15}
\end{equation}
respectively. The reconstructed matrix $V$ is shown in Fig.~\ref{fig4}(a). The obtained similarity in Eq.~(\ref{eq11}) between $\mathcal{V}_{M}$ and $\mathcal{V}_{V}$ is $S_{MV}=0.937\pm0.003$, which demonstrates that $V$ describes our directionally-unbiased $3 \times 3$ linear optical multiport well.
	
	
In order to confirm that our implemented multiport is a directionally-unbiased $3 \times 3$, we independently reconstruct a transfer matrix $W=U_{B}\varPhi U_{F}$. To obtain $W$, we independently reconstruct $U_{F}$ and $U_{B}$ by measuring the HOM visibility matrix and the amplitude distribution, respectively. We performed this measurement by removing mirror units and connecting the outputs of the tritter into the SNSPDs. The obtained HOM visibility matrix and the amplitude distribution data are presented in Appendix A. With these data, the experimentally reconstructed $U_{F}$ and the corresponding standard deviation $\sigma_{F}$ are
\begin{equation}
U_{F}=\left(\begin{array}{ccc} 0.579 & 0.525 & 0.612\\0.626 & 0.598 \, e^{i \, 0.671\pi} & 0.528 \, e^{-i \, 0.607\pi}\\0.531 & 0.583 \, e^{-i \, 0.711\pi} & 0.594 \, e^{i \, 0.706\pi} \end{array}\right),
\label{eq16}
\end{equation}
and
\begin{equation}
\sigma_{F}=\left(\begin{array}{ccc} 0.034 & 0.023 & 0.023 \\ 0.019 & 0.026 \, e^{i \, 0.008\pi} & 0.025 \, e^{i \, 0.008\pi} \\ 0.026 & 0.027 \, e^{i \, 0.007\pi} & 0.028 \, e^{i \, 0.007\pi} \end{array}\right),
\label{eq17}
\end{equation}
respectively. Likewise, the experimentally reconstructed $U_{B}$ and the corresponding standard deviation $\sigma_{B}$ are given by
\begin{equation}
U_{B}=\left(\begin{array}{ccc} 0.567 & 0.602 & 0.552\\0.549 & 0.583 \, e^{i \, 0.654\pi} & 0.590 \, e^{-i \, 0.692\pi}\\0.603 & 0.541 \, e^{-i \, 0.648\pi} & 0.597 \, e^{i \, 0.670\pi} \end{array}\right),
\label{eq18}
\end{equation}
and
\begin{equation}
\sigma_{B}=\left(\begin{array}{ccc} 0.026 & 0.028 & 0.030 \\ 0.032 & 0.033 \, e^{i \, 0.008\pi} & 0.027 \, e^{i \, 0.007\pi} \\ 0.030 & 0.029 \, e^{i \, 0.006\pi} & 0.025 \, e^{i \, 0.005\pi} \end{array}\right),
\label{eq19}
\end{equation}
respectively.

By substituting reconstructed $U_F$ and $U_{B}$ to Eq.~(\ref{eq4}), we numerically obtain the $\varPhi$ by minimizing the $RMS2=\Sigma(\mathcal{V}_{W}-\mathcal{V}_{M})^{2}$ by varying the $\phi_1$ and $\phi_2$, here $\mathcal{V}_{W}$ denotes the calculated visibility matrix from $W$. By combining the obtained phase parameters $[\phi_{1}, \phi_{2}]=[(0.383\pm0.014)\pi, (-0.596\pm0.010)\pi]$ and the transfer matrices $U_{B}$ and $U_{F}$, we obtain the transfer matrix of the implemented $3 \times 3$ directionally-unbiased multiport $W_{0}$. Then, we can obtain the corresponding equivalent real-bordered matrix $W$ and standard deviation, which are given by
		
	
\begin{equation}
W=\left(\begin{array}{ccc} 0.383 & 0.314 & 0.903\\0.314 & 0.888 \, e^{-i \, 0.407\pi} & 0.316 \, e^{-i \, 0.722\pi}\\0.903 & 0.265 \, e^{-i \, 0.662\pi} & 0.203 \, e^{i \, 0.920\pi} \end{array}\right),
\label{eq20}
\end{equation}
and
\begin{equation}
\sigma_{W}=\left(\begin{array}{ccc} 0.014 & 0.012 & 0.008 \\ 0.011 & 0.008 \, e^{i \, 0.024\pi} & 0.015 \, e^{i \, 0.020\pi} \\ 0.009 & 0.015 \, e^{i \, 0.022\pi} & 0.010 \, e^{i \, 0.297\pi} \end{array}\right),
\label{eq21}
\end{equation}
respectively, where the real-bordered matrix denotes the matrix form described on Eq. (\ref{eq5}). The detailed information on how we obtain the real-bordered matrix is provided in Appendix B. The similarity between $\mathcal{V}_M$ and $\mathcal{V}_{W}$ is $S_{MW}=0.972\pm0.007$ and $W$ is graphically shown in Fig.~\ref{fig4}(b). 
	
	In order to compare the reconstructed transfer matrices $V$ and $W$, we use fidelity $F$ between two matrices, which  is defined by~\cite{Polino2019}
	\begin{equation}
	F(V, W)=\dfrac{|Tr[V^{\dagger}W]|}{3},
	\label{eq22}
	\end{equation}
where $0\le F\le1$. High fidelity means matrices are similar with each other. The obtained fidelity is $F(V, W)=0.971\pm0.005$, meaning that these two transfer matrices are highly similar to each other and verifies Eq.~(\ref{eq4}). 

\section{conclusion}
	
We experimentally demonstrated a $3 \times 3$ directionally-unbiased linear optical multiport consisting of a tritter and mirrors. We reconstructed the transfer matrix from the experimental two-photon HOM visibility and amplitude distributions via two independent reconstruction methods, and confirmed that two methods give the similar results in terms of the fidelity. Comparing to the previous demonstration of a $3 \times 3$ directionally-unbiased multiport based on bulk optics elements~\cite{Osawa2018}, our $3 \times 3$ directionally-unbiased linear optical multiport does not require a long coherence length of the light source. We also note that our experiment is the first demonstration of quantum optical phenomenon using  $3 \times 3$ directionally-unbiased multiport. We emphasize that our scheme can be easily extended to implement a $N \times N$ directionally-unbiased linear optical multiport by replacing a tritter into a $N \times N$ multiport BS.

	
Since a directionally-unbiased multiport can be a single vertex having multiple edges in a graph, we can perform QW simulation on a complex graph networks by combining a number of directionally-unbiased multiports. For example, our $3 \times 3$ directionally-unbiased multiport can construct a hexagonal lattice. Complex graph networks composed of directionally-unbiased multiports can also be used for simulating various Hamiltonians and topological physics~\cite{Simon2017,Simon2017_2,Simon2020, Simon2018}. Moreover, our directionally-unbiased multiport can be imprinted on an integrated photonic circuits~\cite{Peruzzo2011} so that the phases of the multiport can be even more stabilized and the relative phases can be tunable electrically. We believe that our results can provide a practical platform to perform quantum simulations on complex graph networks.

\subsection*{Appendix A: Experimental results of a forward and backward propagating tritter}
	
	For direct reconstruction of a single tritter transfer matrix $U_{F}$ and $U_{B}$, we experimentally obtained the HOM visibility and amplitude distributions of our tritter in both forward and reverse directions. Reverse propagation of the tritter is realized by exchanging the input and output ports of the tritter device. Experimental results are  
	
	\begin{equation}
	\mathcal{V}_{F}=\left(\begin{array}{ccc} 0.499\pm0.038 & 0.617\pm0.021 & 0.323\pm0.025\\0.213\pm0.028 & 0.501\pm0.012 & 0.384\pm0.014\\0.454\pm0.047 & 0.155\pm0.023 & 0.400\pm0.025 \end{array}\right),
	\label{eq23}
	\end{equation}
	and 
	\begin{equation}
	|u_{F}|^{2}=\left(\begin{array}{ccc} 0.348\pm0.017 & 0.305\pm0.016 & 0.347\pm0.017\\0.362\pm0.017 & 0.331\pm0.016 & 0.307\pm0.015\\0.310\pm0.017 & 0.361\pm0.015 & 0.329\pm0.016 \end{array}\right),
	\label{eq24}
	\end{equation}
	and
	\begin{equation}
	\mathcal{V}_{B}=\left(\begin{array}{ccc} 0.442\pm0.035 & 0.417\pm0.014 & 0.556\pm0.024\\0.574\pm0.028 & 0.511\pm0.011 & 0.427\pm0.021\\0.450\pm0.040 & 0.513\pm0.014 & 0.478\pm0.027 \end{array}\right),
	\label{eq25}
	\end{equation}
	and
	\begin{equation}
	|u_{B}|^{2}=\left(\begin{array}{ccc} 0.340\pm0.014 & 0.342\pm0.014 & 0.318\pm0.017\\0.312\pm0.015 & 0.331\pm0.015 & 0.357\pm0.015\\0.343\pm0.017 & 0.317\pm0.015 & 0.340\pm0.014 \end{array}\right).
	\label{eq26}
	\end{equation}
	
	\subsection*{Appendix B: Phase parameters to obtain the equivalent real-bordered unitary matrix}
	
	The relationship between two equivalent matrices $W$ and $W_{0}$ is presented as
	\begin{equation}
	W=\left(\begin{array}{ccc} \textit{e}^{i\phi_{a}} & 0 & 0\\0 & \textit{e}^{i\phi_{b}} & 0\\0 & 0 & \textit{e}^{i\phi_{c}} \end{array}\right)W_{0}\left(\begin{array}{ccc} 1 & 0 & 0\\0 & \textit{e}^{i\phi_{d}} & 0\\0 & 0 & \textit{e}^{i\phi_{e}} \end{array}\right).
	\label{eq27}
	\end{equation}
	Since both $W$ and $W_{0}$ give the same HOM visibility and amplitude distributions, $W$ is also a transfer matrix of our $3 \times 3$ directionally-unbiased multiport. The phase factor $\phi_{m}$ with $m=a, b, c, d, e$ correspond to the relative phase on the input and output ports of the multiport. Each $\phi_{m}$ is determined by solving first order equations, which satisfy the real-border condition that $W_{k1}$ and $W_{1i}$ are real numbers for all $k$ and $i$.


\section*{Acknowledgements}
This work was supported by the National Research Foundation of Korea (NRF) (2019M3E4A1079777, 2019M3E4A1078660, 2021R1C1C1003625), the Institute for Information and Communications Technology Promotion (IITP) (2020-0-00947, 2020-0-00972), and the KIST research program (2E31021).

\end{document}